\documentclass[journal=jacsat,manuscript=article]{achemso}

\usepackage[version=3]{mhchem} 
\usepackage{booktabs}
\usepackage{graphicx}
\usepackage{amssymb}
\usepackage{setspace}
\usepackage[utf8]{inputenc}
\usepackage[T1]{fontenc}
\usepackage[detect-weight, separate-uncertainty=true]{siunitx}
\usepackage{relsize}
\usepackage[nomain, acronym]{glossaries}
\usepackage[justification=centering]{caption}
\usepackage{natbib}
\setacronymstyle{long-sm-short}
\glsdisablehyper

\makeatletter
\newcommand\etc{etc\@ifnextchar.{}{.\@}}
\newcommand\etal{\emph{et al}\@ifnextchar.{}{.\@}}
\makeatother

\newacronym{CMOS}{CMOS}{complementary metal-oxide-semiconductor}
\newacronym{SDM}{SDM}{spatial division multiplexing}
\newacronym{WDM}{WDM}{wavelength division multiplexing}
\newacronym{MDM}{MDM}{mode division multiplexer}
\newacronym{CWDM}{CWDM}{coarse wavelength division multiplexer}
\newacronym{HSQ}{HSQ}{hydrogen silsesquioxane}
\newacronym{SOI}{SOI}{silicon-on-insulator}
\newacronym{SiN}{SiN}{silicon nitride}
\newacronym{MZI}{MZI}{Mach-Zehnder interferometers}
\newacronym{MSI}{MSI}{medium-scale integration}
\newacronym{LSI}{LSI}{large-scale integration}
\newacronym{VLSI}{VLSI}{very-large-scale integration}
\newacronym{CPO}{CPO}{co-packaged optics}
\newacronym{SiO2}{SiO\ensuremath{_2}}{silicon dioxide}
\newacronym{SNOI}{SNOI}{silicon nitride-on-insulator}
\newacronym{SR}{SR}{split ratio}
\newacronym{IL}{IL}{insertion loss}
\newacronym{XT}{XT}{crosstalk}
\newacronym{SPINS}{SPINS}{Inverse Design Software for Nanophotonic Structures}\glsunset{SPINS}
\newacronym{SEM}{SEM}{scanning electron microscopy}
\newacronym{MEMS}{MEMS}{electrothermal micro-electromechanical systems}
\newacronym{EBL}{EBL}{electron-beam lithography}
\newacronym{OSA}{OSA}{optical spectrum analyzer}
\newacronym{FDTD}{FDTD}{finite-difference time-domain}
\newacronym{PECVD}{PECVD}{plasma-enhanced chemical vapor deposition}
\newacronym{RIE}{RIE}{reactive ion etching}
\newacronym{ANT}{ANT}{Applied Nanotools}
\newacronym{MPW}{MPW}{multi-project wafer}
\newacronym{SMF}{SMF}{single-mode fiber}
\newacronym{PIC}{PIC}{photonic integrated circuit}
\newacronym{BOX}{BOX}{buried oxide}
\newacronym{COVID}{COVID}{}\glsunset{COVID}
\newacronym{PBS}{PBS}{polarization beam splitter}
\newacronym{DFM}{DFM}{design for the manufacturing}
\newacronym{PER}{PER}{polarization extinction ratio}
\author{Julian L.\ Pita Ruiz}
\email{julian-leonel.pita-ruiz@etsmtl.ca}
\affiliation[ETS]
{Department of Electrical Engineering, École de Technologie Supérieure (ÉTS), Montreal, QC H3C 1K3, Canada}

\author{Narges Dalvand}
\affiliation[ETS]
{Department of Electrical Engineering, École de Technologie Supérieure (ÉTS), Montreal, QC H3C 1K3, Canada}

\author{Michaël Ménard}
\affiliation[ETS]
{Department of Electrical Engineering, École de Technologie Supérieure (ÉTS), Montreal, QC H3C 1K3, Canada}

\title[SiN devices]
  {Integrated Silicon Nitride Devices via Inverse Design}

\begin{document}


\begin{abstract}
    
Integrated photonic devices made of \gls{SiN}, which can be integrated with silicon-on-insulator and III-V platforms, are expected to drive the expansion of silicon photonics technology.
However, the relatively low refractive index contrast of \gls{SiN} is often considered a limitation for creating compact and efficient devices.
Here, we present three freeform \gls{SiN} devices—a coarse wavelength-division multiplexer, a five-mode mode-division multiplexer, and a polarization beam splitter—while systematically benchmarking both the design capability and the fabrication repeatability and robustness of inverse-designed components.
We demonstrate up to a \SI{1200}{\times} reduction in footprint while maintaining relatively large minimum feature sizes of up to \SI{160}{\nano\meter}, showing that inverse-designed \gls{SiN} devices can be as compact as their silicon counterparts.
These results enable high-density integration in \gls{SiN} photonics and pave the way for multidimensional data transmission and quantum applications, as the inverse design technique can be applied to different \gls{SiN} thicknesses and is potentially extendable to other low- and mid-index platforms.

\end{abstract}

\section*{Introduction}

\Gls{SiN} has enabled a wide range of applications, including sensing \cite{harame1987ion,woias1998slow, antonacci2020ultra}, optical communications \cite{butler2021demonstration}, interconnects \cite{yi2024asymmetric, he2024broadband, brunetti2023silicon}, spectroscopy \cite{dhakal2014evanescent, dhakal2014silicon, zhao2018stimulated}, LIDAR \cite{xu2022fully, lukashchuk2024photonic}, and quantum information processing \cite{taballione20198, prokhodtsov2020silicon, senichev2022silicon}, due to its broad transparency window and accessibility. Its impact also results from the ultra-low waveguide propagation losses it can provide \cite{blumenthal2018silicon}, with losses below \SI{0.1}{\decibel\per\meter}, facilitating the management of high optical power and enabling nonlinear applications such as optical frequency comb generation \cite{kim2017dispersion, guo2018mid}.
The small thermo-optic coefficient of \gls{SiN} and \gls{SiO2} results in a low wavelength dependence on temperature \cite{xiang2022silicon}. This property benefits passive devices but presents challenges for tunable components. Additionally, the relatively low refractive index contrast between \gls{SiN} and \gls{SiO2} minimizes phase errors in devices such as interferometers, enhancing robustness against fabrication imperfections \cite{baets2016silicon}. However, this advantage comes at the expense of larger device footprints, which limits the number of components that can be integrated per chip. For instance, in a platform where the \gls{SiN} is \SI{400}{nm} thick, a \SI{850}{nm} wide waveguide should be routed with a minimum bend radius of \SI{50}{\um} to achieve negligible losses \cite{ruiz2025inverse}.

Silicon photonics is transitioning from the era of medium-scale integration (MSI; 10–500 components) to that of large-scale integration (LSI; 500–10000 components), driven in part by the demands of photonic quantum computing \cite{shekhar2024roadmapping}. The rise of co-packaged optics (CPO) demands LSI photonics—and eventually very-large-scale integration (VLSI) chips with more than 10,000 components—offering high bandwidth and connectivity \cite{psiquantum2025manufacturable}, alongside high-performance electronic chips within the same package. This level of integration is essential for enabling the next generation of high-performance computing systems, including those driving artificial intelligence \cite{ahmed2025universal}. For \gls{SiN} to play a key role in \gls{CPO} and VLSI photonic circuits, substantial reductions in component sizes, waveguide routing bends, and input/output coupler dimensions are required.

Spatial and mode multiplexers, along with polarization management devices, are essential components in many applications, especially in optical communications \cite{caut2023channel, sabri2024high}. Recently, various low-loss multiplexers and polarization splitters implemented on a silicon nitride platform have been reported \cite{yang2014silicon, tu2023400, zheng2023high, bhandari2020highly, guerber2018broadband, vanani2024broadband}. However, most demonstrations require a substantial footprint, and attempts to miniaturize these devices often introduce additional fabrication complexity, such as deep etch steps or the integration of additional material layers \cite{feng2014three, sun2017realization, gallacher2022silicon}.
\Gls{SiN} multiplexers can be implemented using gratings \cite{sciancalepore2018band, shi2013ultra}, cascaded \gls{MZI} \cite{cheung2020silicon, xie2018silicon}, ring resonators \cite{dorin2014two}, and directional couplers \cite{yang2014silicon}. Gao et al. demonstrated a \gls{CWDM} for the O-band with a footprint of \num{1} $\times$ \SI{0.6}{mm^2}, achieving a minimum loss of \SI{1.55}{dB} \cite{gao2017silicon}. More recently, a three-mode \gls{MDM} with a length of \SI{1.455}{mm} operating at visible wavelengths was demonstrated, achieving an insertion loss of \SI{0.8}{dB} for the highest-order mode \cite{shah2024mode}. Kudalippalliyalil et al. presented a compact polarization splitter measuring \num{80} $\times$ \SI{13}{\um^2} on a \SI{400}{\nm} thick SiN platform, achieving an efficiency of up to \SI{-0.7}{dB} and a crosstalk level below \SI{-18.0}{dB} \cite{kudalippalliyalil2020low}.

In this work, we demonstrate the potential of the inverse design technique to improve component integration density on photonic platforms with moderate refractive index contrast by presenting  three ultra-compact freeform devices—a \gls{CWDM}, a five-mode \gls{MDM}, and a \gls{PBS}—fabricated on a single-etch \gls{SiN} platform. The four-channel \gls{CWDM}, with a footprint of \num{24}$\times$\SI{24}{\um^2}, achieves a maximum measured efficiency of \SI{-1.0}{dB} across all channels. The five-mode \gls{MDM}, with a compact \SI{16}{\um} length, achieves an average maximum efficiency of \SI{-1.0}{dB} for the TE$_{00}$ mode and \SI{-2.3}{dB} for the TE$_{04}$ mode, with crosstalk below \SI{-12.0}{dB}. The polarization beam splitter, measuring \num{24}$\times$\SI{12}{\um^2}, exhibits an average efficiency of \SI{-0.8}{dB} for the TE mode and of  \SI{-0.9}{dB} for the TM mode across the entire C-band, with crosstalk as low as \SI{-18.0}{dB}. These devices feature the smallest reported footprints among similar demonstrated devices, significantly enhancing integration density while maintaining high performance.

\section{Results}

All proposed devices were implemented using a fully etched \gls{SiN} core with a thickness of \SI{400}{\nano\meter}, surrounded by \gls{SiO2}. The refractive indices at \SI{1550}{\nano\meter} were \num{1.997} for \gls{SiN} and \num{1.441} for \gls{SiO2}. We employed the Python-based wrapper LumOpt, which enables density-based topology optimization to design the structures while incorporating fabrication constraints, including a minimum feature size and a smoothing filter with a radius of up to \SI{160}{\nano\meter}. The devices were fabricated using \gls{EBL} and characterized using a wafer-level automated test station (see Methods).

\subsection{\gls{CWDM}}

\begin{figure}[b!]
\centering
\includegraphics[width=\linewidth]{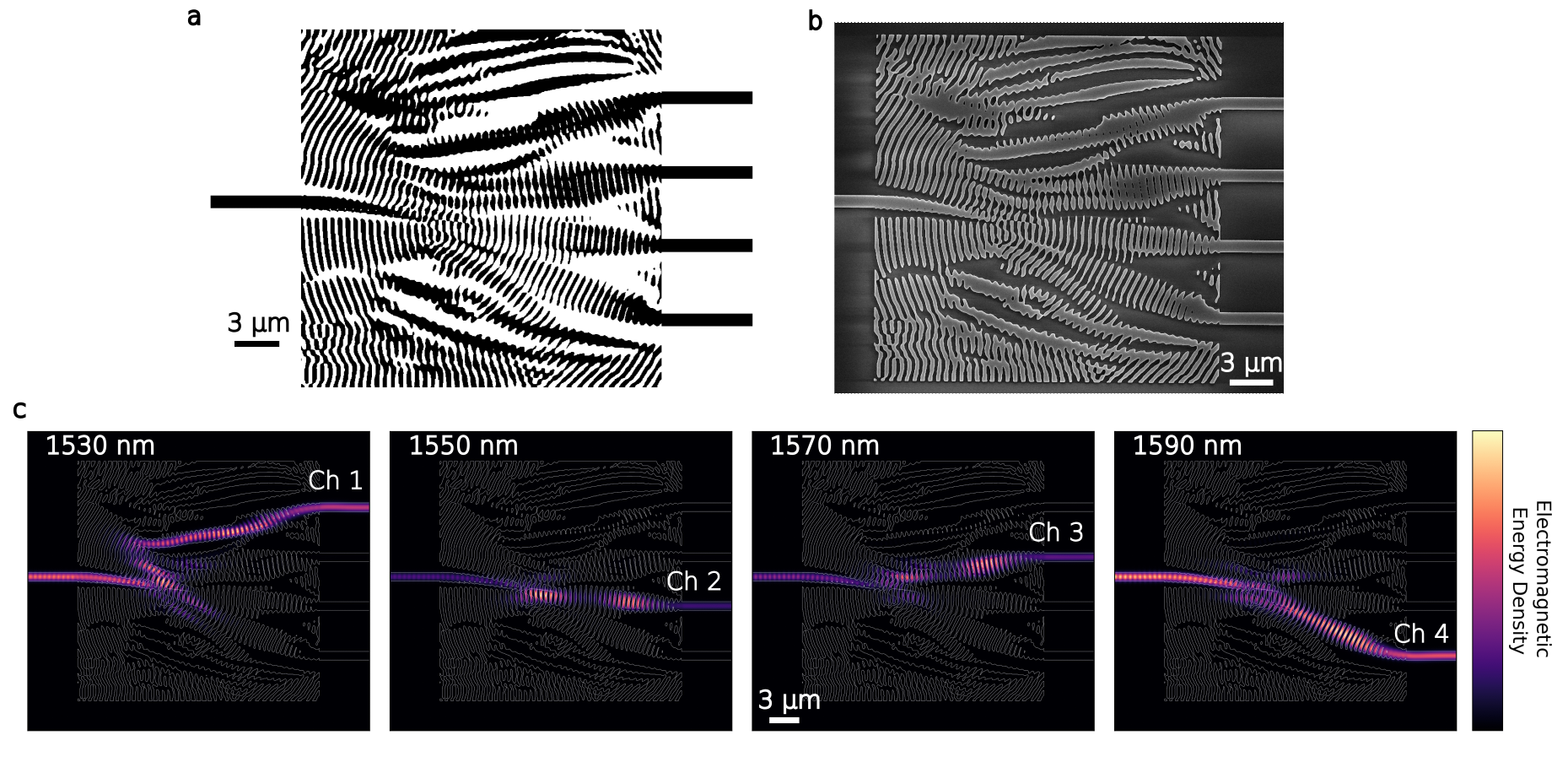}
\caption{Four-channel \gls{CWDM} multiplexer. (a) Device design, where the black areas represent silicon nitride and the white ones represent silicon dioxide. (b) SEM image of the fabricated device, with a total footprint of \num{24} $\times$ \SI{24}{\um^2}. (c) Simulated electromagnetic energy density at the central operating wavelengths of each channel.}
\label{fig:1}
\end{figure}

The compact \gls{CWDM} was designed using the fundamental TE mode of the input waveguide as the input mode for the inverse design process, while the fundamental TE modes of the four output waveguides were used as the output modes. The waveguide width was chosen as \SI{850}{nm} to ensure single-mode operation. The target central wavelengths were \SI{1530}{nm}, \SI{1550}{nm}, \SI{1570}{nm}, and \SI{1590}{nm}, and the optimization was carried out using five wavelengths within a \SI{10}{nm} bandwidth around each central wavelength. Figure~\ref{fig:1} presents the final device design, a \gls{SEM} image of the fabricated \num{24}$\times$\SI{24}{\um^2} device, and the simulated electromagnetic energy density at each channel central wavelength. At \SI{1550}{nm}, the light undergoes a directional change before propagating through the upper channel. In contrast, for the other channels, the light follows a relatively smooth path towards their respective output (see Supplementary Movie).

\begin{figure}[b!]
\centering
\includegraphics[scale=1.1]{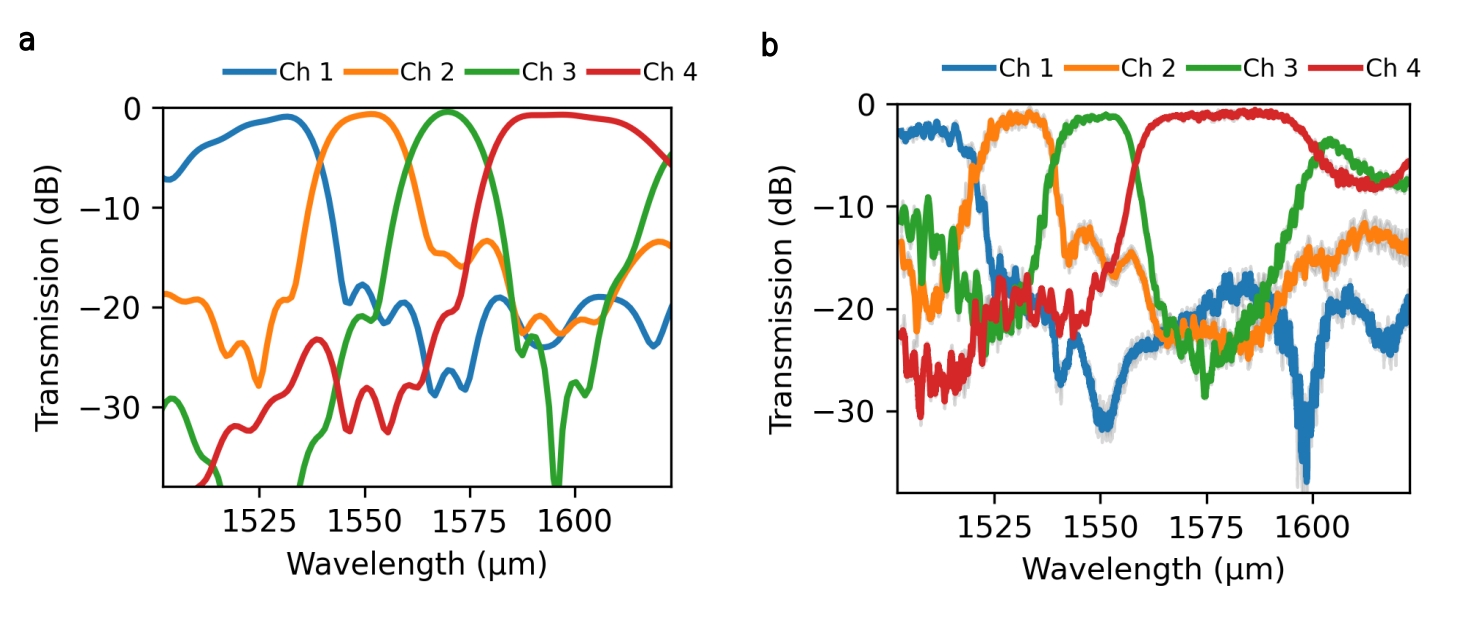}
\caption{Simulated and measured transmission efficiency of the four-channel \gls{CWDM}. (a) Simulated transmission calculated using a 3D \gls{FDTD} algorithm. (b) Measured transmission of three copies of the CDWM across the four channels. The solid lines indicate the average of the three devices, and the grey shaded region is bounded by the minimum and maximum measured transmission.}
\label{fig:2}
\end{figure}

The simulated and measured transmission results are shown in Figure~\ref{fig:2}. The simulated transmission and crosstalk are \SI{-0.9}{dB} and \SI{-19.3}{dB} at \SI{1530}{nm}, \SI{-0.7}{dB} and \SI{-18}{dB} at \SI{1550}{nm}, \SI{-0.5}{dB} and \SI{-15.2}{dB} at \SI{1570}{nm}, and \SI{-0.7}{dB} and \SI{-20.8}{dB} at \SI{1590}{nm}. Simulations also indicate that backreflection into the input waveguide remains consistently below \SI{-16.9}{dB} at central wavelengths. Measurements from three identically fabricated devices, shown in Figure~\ref{fig:2}b, demonstrate highly repeatable performance. The average measured transmission was \SI{-1.7}{dB} at \SI{1511.3}{nm}, \SI{-1.3}{dB} at \SI{1531.3}{nm}, \SI{-1.0}{dB} at \SI{1551.3}{nm}, and \SI{-1.0}{dB} at \SI{1571.3}{nm}. At these wavelengths, the measured crosstalk remained below \SI{-13.3}{dB}. The discrepancies between the simulated and measured results are primarily attributed to over etching during the fabrication process (see Supplementary Figure 4).

\subsection{MDM}

\begin{figure*}[b!]
\includegraphics[width=\linewidth]{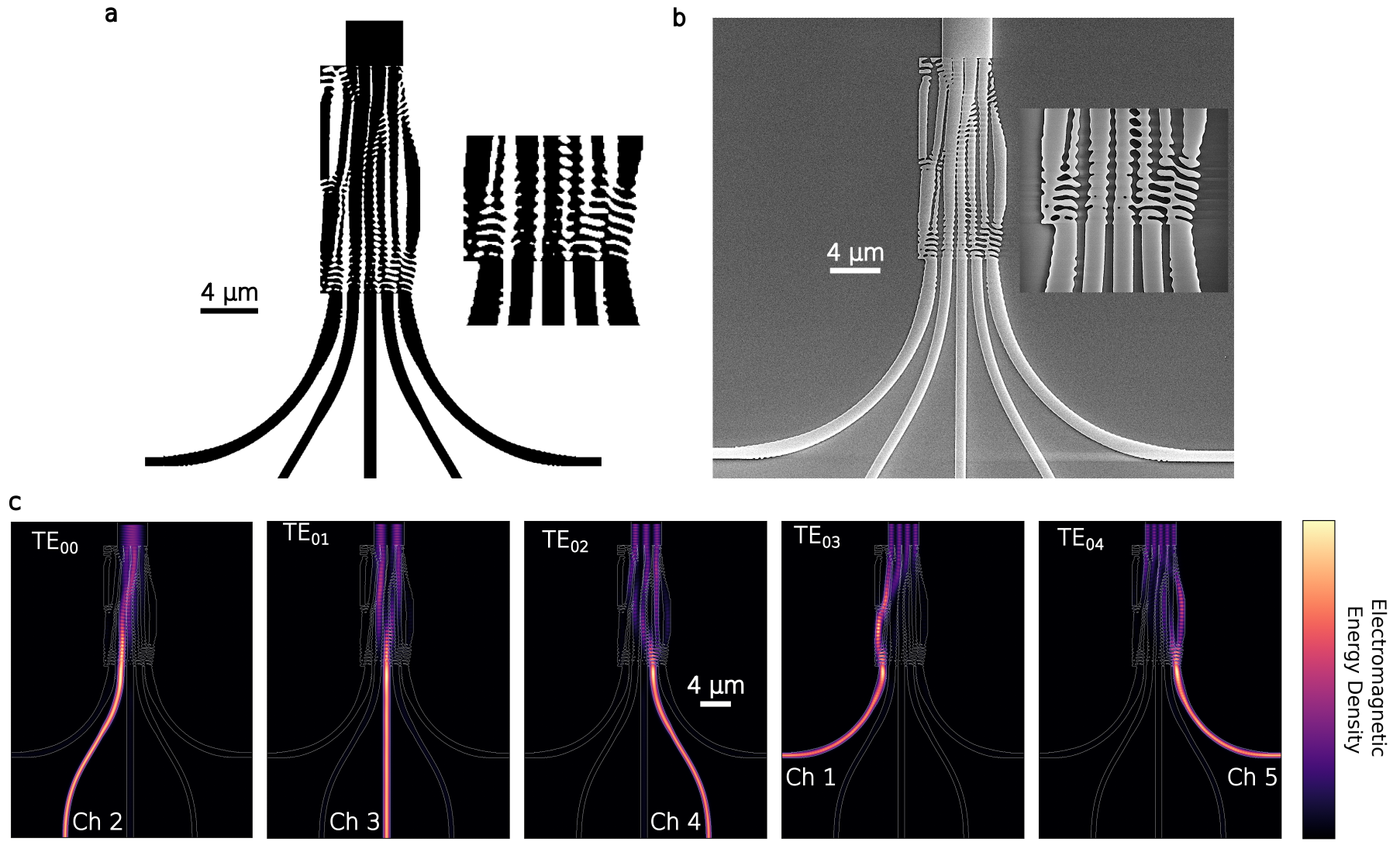}
\caption{Five-mode \gls{MDM} multiplexer. (a) Device design and routing network, where the black areas represent silicon nitride and the white ones represent silicon dioxide. (b) SEM image of the fabricated device with a total footprint of \num{16}$\times$\SI{7}{\um^2}, including an inset showing the input waveguides. (c) Simulated electromagnetic energy density for each channel at \SI{1550}{nm}.}
\label{fig:3}
\end{figure*}

A compact, inverse-designed five-mode \gls{MDM} multiplexer was optimized using the fundamental TE mode of five \SI{850}{nm}-wide input waveguides,  spaced \SI{1.2}{\um} apart, as the input modes, while the first five TE modes of a \SI{4}{\um}-wide output waveguide served as the output modes. Broadband optimization, spanning the entire C- and L-bands, was performed over a rectangular design region of \num{16}$\times$\SI{7}{\um^2}, with a minimum feature size and a smoothing filter radius of \SI{120}{nm} applied during the optimization process. Figure~\ref{fig:3} presents the final device design, including the routing network, a \gls{SEM} image of the fabricated structure (with an inset showing the region near the input waveguides), and the simulated electromagnetic energy density at \SI{1550}{nm} for each mode. The five-mode \gls{MDM} multiplexer routes the fundamental TE$_{00}$ mode of the input single-mode waveguides to the TE$_{00}$ (channel 2), TE$_{01}$ (channel 3), TE$_{02}$ (channel 4), TE$_{03}$ (channel 1), and TE$_{04}$ (channel 5) modes of the output multimode waveguide. Additionally, \SI{16}{\um} radius freeform bends were used to route channels 1 and 5 \cite{ruiz2025inverse}, minimizing crosstalk. Furthermore, the five-mode \gls{MDM} can be used to implement an \gls{MEMS} inter-chip modal switch \cite{ruiz20241} if instead of using five fixed input waveguides only one input, mounted on a \gls{MEMS} actuator, is used to select which higher order mode is excited. 

\begin{figure*}[b!]
\centering
\includegraphics[width = \linewidth]{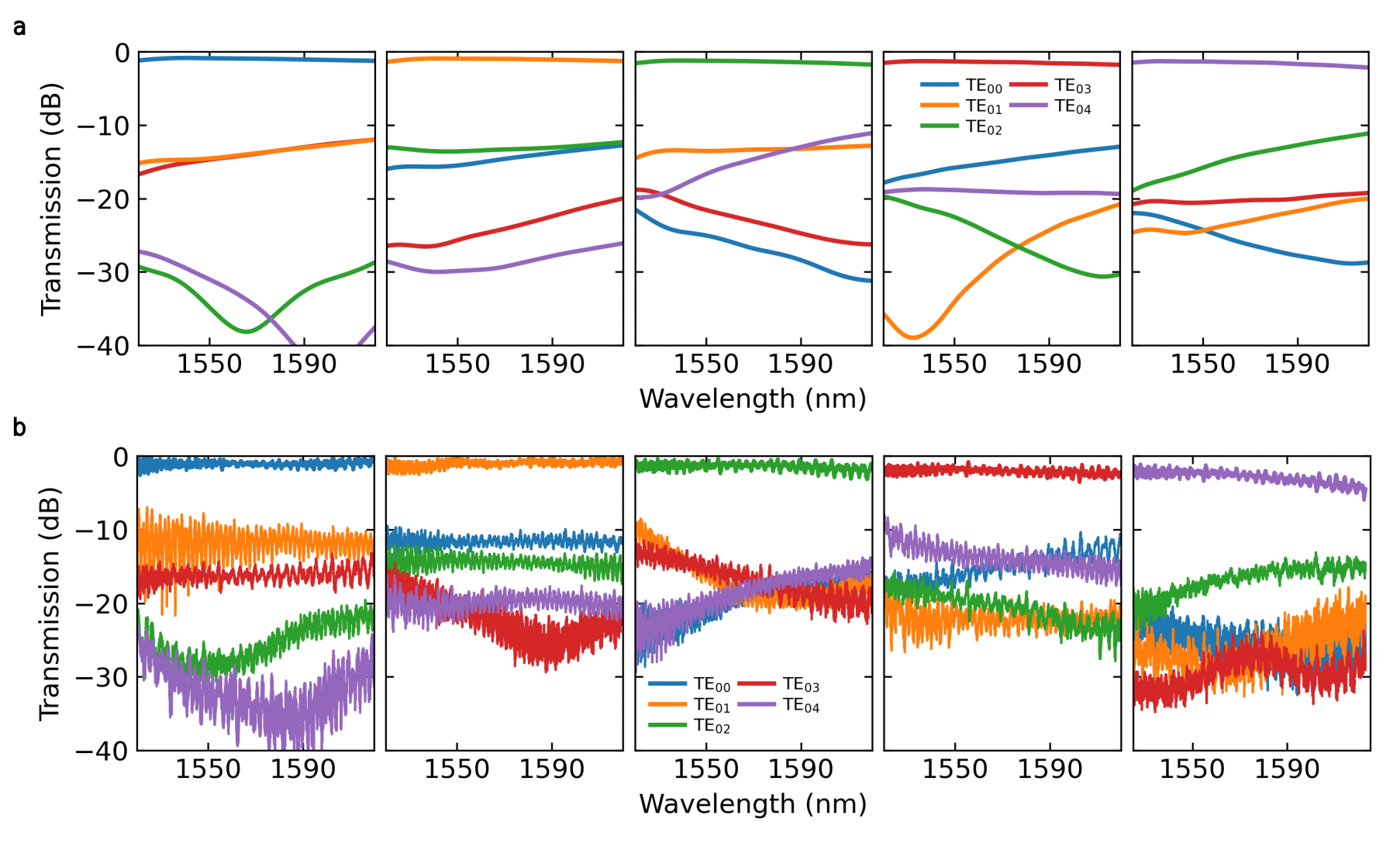}
\caption{Simulated and measured transmission efficiency of the five-mode \gls{MDM}. (a) Simulated transmission calculated using 3D FDTD. (b) Measured transmission of the fabricated device for all five channels.}
\label{fig:4}
\end{figure*}

The simulated and measured transmission results are shown in Figure~\ref{fig:4}. The average simulated transmission efficiency across the entire C-band is \SI{-0.9}{dB} for TE$_{00}$, \SI{-0.9}{dB} for TE$_{01}$, \SI{-1.2}{dB} for TE$_{02}$, \SI{-1.3}{dB} for TE$_{03}$, and \SI{-1.3}{dB} for the TE$_{04}$ mode, respectively. The 3-dB bandwidth exceeds \SI{120}{nm}, and the average crosstalk across these wavelengths remains below \SI{-13.5}{dB} for all channels. The measured average efficiency in the C-band is \SI{-1.0}{dB} for TE$_{00}$, \SI{-1.1}{dB} for TE$_{01}$, \SI{-1.3}{dB} for TE$_{02}$, \SI{-1.9}{dB} for TE$_{03}$, and \SI{-2.3}{dB} for TE$_{04}$, with crosstalk values of \SI{-12.0}{dB}, \SI{-11.7}{dB}, \SI{-14.9}{dB}, \SI{-13.0}{dB}, and \SI{-18.4}{dB}, respectively. Furthermore, the transmission spectra is flat across the entire C-band. The five-mode \gls{MDM} multiplexer, including its routing network, maintains efficiencies comparable to those of its silicon counterparts while preserving its ultra-compact footprint \cite{sun2025edge}. Modal crosstalk can be reduced by increasing the channel spacing in the five-mode \gls{MDM} multiplexer; however, this would also increase the switching time when used as part of a \gls{MEMS} switch.

\subsection{PBS}

\begin{figure*}[t!]
\centering
\includegraphics[scale=1.1]{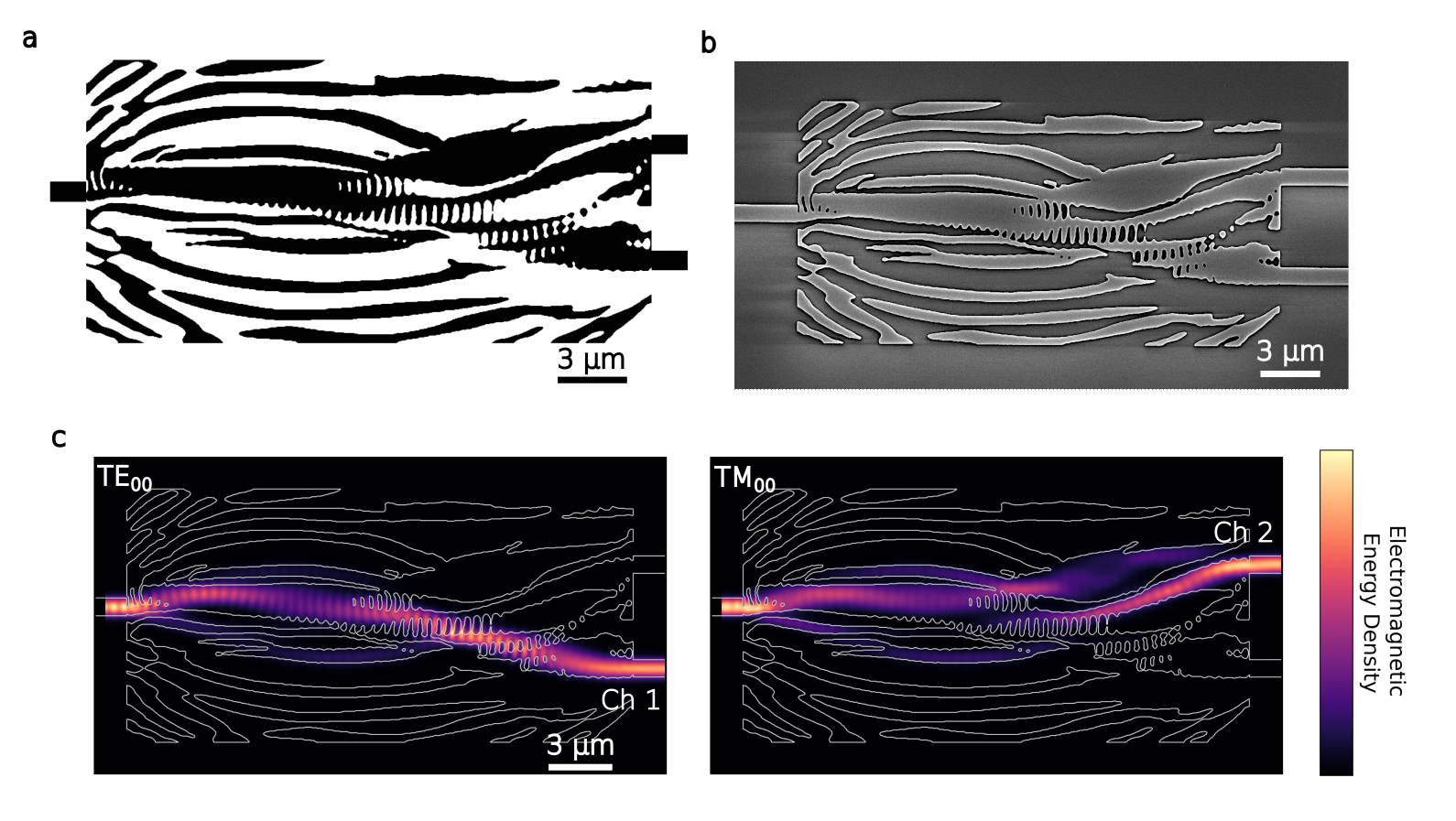}
\caption{Compact \gls{PBS}. (a) Device design, where black areas represent silicon nitride and white areas represent silicon dioxide. (b) SEM image of the fabricated device, which has a total footprint of \num{24}$\times$\SI{12}{\um^2}. (c) Simulated electromagnetic energy density at \SI{1550}{nm} for both the TE and TM polarizations.}
\label{fig:5}
\end{figure*}

Figure~\ref{fig:5} presents the ultra-compact \gls{PBS} design, an SEM image of the fabricated device, and the simulated energy density at \SI{1550}{nm} for both TE and TM polarizations. The \gls{PBS} was designed using the fundamental TE and TM modes of the input waveguide as input modes for the inverse design process, with the TE and TM modes directed to output channels 1 and 2, respectively. The spacing between the output waveguides was set to \SI{4.925}{\um}, and both the input and output waveguide widths were set to \SI{850}{nm}. The optimization was performed within a \num{24}$\times$\SI{12}{\um^2} region, targeting broadband operation spanning from \SI{1520}{nm} to \SI{1630}{nm}, and incorporating a minimum feature size and a smoothing filter with a radius of \SI{160}{nm}.

\begin{figure*}[b!]
\centering
\includegraphics[scale=1.1]{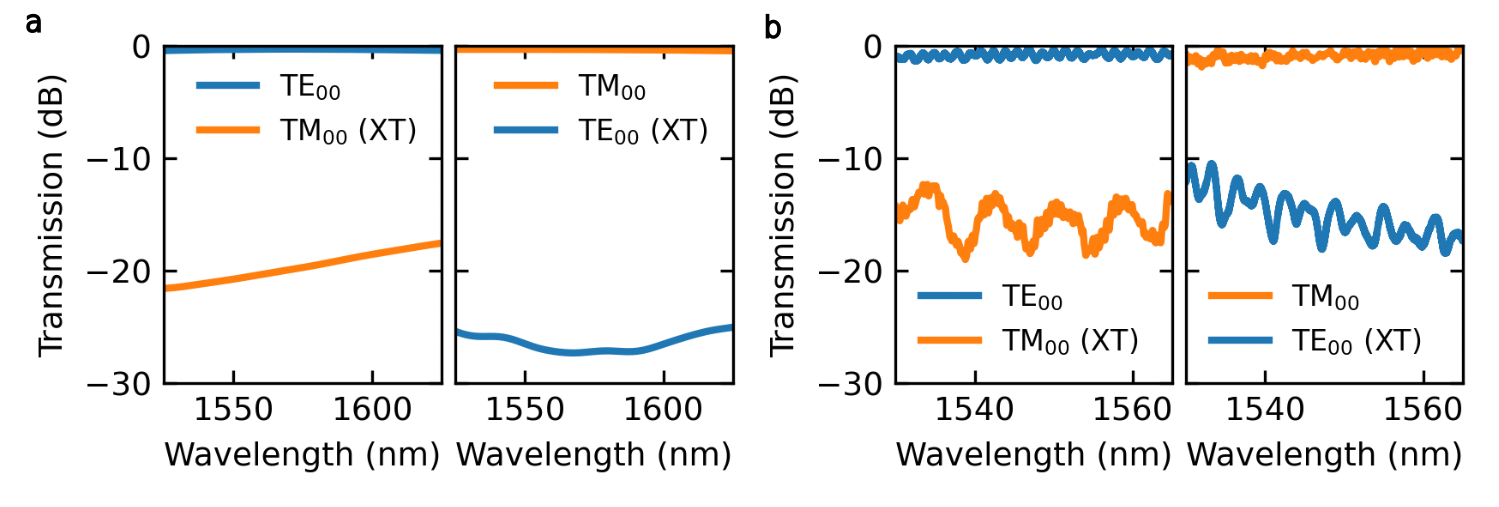}
\caption{Simulated and measured transmission efficiency of the \gls{PBS}. (a) Simulated transmission from \SI{1525}{nm} to \SI{1625}{nm}. (b) Measured transmission in the C-band for the fabricated device for both the TE and TM polarizations. In both cases, the right subplot represents the transmission of channel 1, while the left subplot represents the transmission of channel 2.}
\label{fig:6}
\end{figure*}

The simulated and measured transmission spectra of the compact \gls{PBS} device for channels 1 and 2, corresponding to TE and TM polarizations, respectively, are shown in Figure~\ref{fig:6}. The average simulated TE transmission efficiency to channel 1 is \SI{-0.3}{dB}, while the undesired TM transmission to the same channel averages \SI{-20.1}{dB} across the C-band. Similarly, the average simulated TM transmission efficiency to channel 2 is \SI{-0.3}{dB}, with the undesired TE transmission averaging \SI{-25.8}{dB}. These values indicate a simulated \gls{PER} exceeding \SI{19.8}{dB}. 
The measured average TM transmission efficiency to channel 1 is \SI{-0.9}{dB}, while the average TE transmission efficiency to channel 2 is \SI{-0.8}{dB}. The undesired TM transmission to channel 1 and the undesired TE transmission to channel 2 are \SI{-15.5}{dB} and \SI{-15.3}{dB}, respectively. Measurements were limited to the C-band due to constraints imposed by the TM grating coupler (see Methods). The reduction of the experimental \gls{PER} to approximately \SI{14.5}{dB} is attributed to fabrication variations, including over etching and under etching (see Supplementary Figure 5).

Although it may be desirable, most inverse-designed photonic structures do not exhibit periodic, regular, or modular patterns and therefore cannot be easily interpreted or modeled using traditional physical models. However, the devices presented in this work—including the four-channel \gls{CWDM}, the five-mode \gls{MDM}, and the \gls{PBS}—can be understood as systems that precisely engineer light–matter interactions at the subwavelength scale. Through their non-intuitive, freeform geometries, these devices spatially tailor the effective refractive index to produce highly specific patterns of constructive and destructive interferences, each optimized for particular wavelengths, spatial modes, or polarizations. Thus, the dispersion required to separate wavelengths, polarizations or to manipulate a wavefront to convert it into a different spatial mode is achieved by engineering the refractive index of the subwavelength pixels. Therefore, this approach provides capabilities that go far beyond conventional photonic components.

\section{Discussion}

The inverse design technique has been increasingly utilized in both academia and industry to enhance component density on \gls{SOI} platforms. Despite the relatively high refractive index contrast of these platforms, many passive and active devices have been successfully miniaturized using this technique \cite{molesky2018inverse}. However, inverse design has been less explored for platforms with medium to low index contrast, such as silicon nitride, with only a few exceptions, including reflectors \cite{pita2024inverse}, TM polarizers \cite{pita2023silicon}, mode converters \cite{zhou2023tunable}, and beam splitters \cite{song2024ultracompact}. This is primarily due to the significantly higher computational effort required for large optimization regions.

It is important for inverse-designed devices to demonstrate good fabrication repeatability and to remain robust against fabrication variations while maintaining high performance. Our freeform four-channel \num{24}$\times$\SI{24}{\um^2} \gls{SiN} \gls{CWDM} achieves high transmission efficiency, ranging from \SI{-1.0}{dB} to \SI{-1.7}{dB} at the central wavelength of each channel. These results are comparable to those of a freeform four-channel \num{14}$\times$\SI{16}{\um^2} silicon \gls{CWDM} device, which demonstrates efficiency levels between \SI{-2.0}{dB} and \SI{-3.3}{dB} \cite{cheung2024inverse}. Additionally, the unique five-mode freeform \gls{SiN} \gls{MDM} device has a remarkably  small footprint of \num{16}$\times$\SI{7}{\um^2}, compared to \gls{SiN} phase-matching multiplexers, which typically occupy areas above \SI{1}{mm^2}, achieving transmission efficiencies from \SI{-1.0}{dB} to \SI{-2.3}{dB}. This footprint is comparable to the only freeform five-mode \gls{MDM} in silicon presented in the literature, which has a footprint of \num{6}$\times$\SI{10}{\um^2} and achieves a transmission efficiency range of \SI{-2.4}{dB} to \SI{-3.9}{dB} \cite{sun2025edge}. Our inverse-designed \gls{SiN} multiplexer has a size and achieves performance on par or better than its silicon counterparts, despite the lower index contrast.
While our \gls{CWDM} design achieves high efficiency and good fabrication repeatability, the central wavelengths shifted toward lower values due to over etching during fabrication. Fabrication variations also affect crosstalk more than transmission efficiency in broadband devices—including the \gls{MDM} and \gls{PBS}—especially in the case of the \gls{PBS}. These fabrication-related issues can be mitigated using deep learning and fabrication-aware software after the design process \cite{gostimirovic2023improving}, ensuring that miniaturization does not compromise any performance metrics.

In summary, we demonstrated experimentally three ultra-compact freeform silicon nitride devices—a four-channel \gls{CWDM}, a five-mode \gls{MDM}, and a \gls{PBS}—using the inverse design technique. These devices achieve performance levels that have not been demonstrated before in such small \gls{SiN} structures. For instance, the \gls{CWDM} device demonstrates a size reduction of more than \SI{1200}{\times} compared to traditional cascaded \gls{MZI} silicon nitride designs \cite{gao2017silicon}. Our results show that applying inverse design techniques to the silicon nitride platform can enable the development of highly compact devices while preserving superior passive performance, paving the way for large-scale integration.

\section{Methods}

\subsection{Optimization}

To design the devices, we use the Python-based wrapper LumOpt, which supports continuous density-based topology optimization and employs the Ansys Lumerical 3D \gls{FDTD} algorithm to solve the direct and adjoint problems for sensitivity analysis. The optimization is driven by gradient-based algorithms from SciPy.

The topology optimization workflow begins with an initialization step, in which the optimization volume and the initial core and cladding refractive indices are defined. LumOpt uses a rectilinear optimization grid that is fully aligned with the 3D \gls{FDTD} simulation grid used for both the direct and adjoint solves. For each device, we evaluated three initial conditions: the design region was uniformly initialized with either silicon nitride, silicon dioxide, or a refractive index equal to the average of the two. For the \gls{CWDM} and \gls{PBS} devices, the best results were obtained with the average-index initialization, whereas for the \gls{MDM} device, initializing with silicon nitride yielded superior performance.

In the first phase (Greyscale optimization), the material distribution varies continuously between the core and cladding indices. In the second phase (Binarization), the design is forced to take on discrete values corresponding to either the core or cladding material. A convolution filter and a projection function are employed to smooth the design region. Specifically, a circular top-hat kernel with a radius of \num{120} or \SI{160}{\nm} is used to eliminate small holes, sharp corners, and other non-manufacturable features. Simultaneously, an approximation of the Heaviside function transforms the linear mapping of the design parameters into a near step-like function, controlled by the steepness parameter $\beta$. As $\beta \rightarrow \infty$, the geometry becomes increasingly binary. This step may introduce sharp peaks in the figure of merit, which are mitigated through additional optimization iterations to stabilize the design. The final phase (\Gls{DFM}) introduces explicit minimum length-scale geometric constraints to eliminate residual small features and ensure fabrication feasibility. An overview of the topology optimization process is provided in Supplementary Note 4.

The optimization process was performed on a single computer equipped with an Intel Core i9-12900 processor and 64 GB of RAM. The computational time per iteration varied for each device: for the \gls{CWDM}, it ranged from \SI{50}{min} to \SI{5}{h}; for the \gls{MDM}, from \SI{20}{min} to \SI{5.5}{h}; and for the \gls{PBS}, from \SI{15}{min} to \SI{2.5}{h}. The final optimization phase was typically the most time-consuming. Supplementary Figures 1, 2, and 3 illustrate the complete optimization trajectory for each device, along with the material index distribution at representative iterations of each optimization stage.
 
\subsection{Fabrication}

The photonic integration platform used in this work is based on silicon nitride and consists of a \SI{400}{\nano\meter} \gls{SiN} core layer, a \SI{4.5}{\micro\meter} buried oxide layer, a \SI{3}{\micro\meter} top silicon dioxide cladding, and a \SI{525}{\micro\meter}-thick silicon substrate. All devices were fabricated through a multi-project wafer run at \gls{ANT}.

A \gls{HSQ} hard mask was spin-coated onto the SiN layer, and the device patterns were written using \SI{100}{\kilo\electronvolt} \gls{EBL}. The exposed \gls{HSQ} resist was developed using a tetramethylammonium hydroxide-based developer. Pattern transfer into the \gls{SiN} layer was carried out via a single-step anisotropic reactive ion etching process using a $CHF_3/O_2$ chemistry. After etching, a \SI{3}{\micro\meter} top cladding of silicon dioxide was deposited using plasma-enhanced chemical vapor deposition. Although the \gls{EBL} process enables feature sizes as small as \SI{120}{\nano\meter}, a minimum feature size and spacing of \SI{160}{\nano\meter} was enforced for the \gls{CWDM} and \gls{PBS} devices to improve fabrication robustness and ensure compatibility with large-scale \gls{CMOS} foundries using \SI{193}{\nano\meter} deep ultraviolet lithography.

The fabricated waveguides exhibit an average sidewall angle of approximately \ang{83.5}, with measured propagation losses of about \SI{1}{\decibel\per\centi\meter} for the TE mode at \SI{1550}{\nano\meter} under single-mode operation. Further fabrication details are provided in ~\cite{naraine2024moderate}. 

\subsection{Measurement}

The devices were characterized using an EXFO OPAL-EC wafer-level automated test station, which features a motorized high-resolution 4-axis motion system with thermal control, where the chips were placed during measurements. The station also includes a \ang{30} polarization-maintaining 12-fiber array mounted on a hexapod, connected via polarization-maintaining fibers to an EXFO T100S-HP tunable laser and a CTP10 passive optical component testing platform for transmission efficiency measurements. For TM measurements, a polarization rotating fiber was inserted between the laser and the fiber array. A Z-profile scan was conducted on the chip surface to maintain a consistent coupling distance of approximately \SI{20}{\um} throughout all TE measurements. To ensure accurate and repeatable device efficiency measurements, surface coupling optimization was performed using the same automated alignment sequence for all devices. Transmission measurements were then normalized against reference gratings connected in a loopback adjacent to each device, eliminating variations due to coupling and waveguide propagation losses and enabling a direct evaluation of device performance. In addition, to characterize the inverse-designed \gls{MDM}, we used an auxiliary adiabatic directional coupler \gls{MDM} placed back-to-back with the device under test. The two \glspl{MDM} were connected via a \SI{38}{\um}-long, \SI{4}{\um}-wide multimode waveguide. To normalize the measurements and isolate the performance of the inverse-designed \gls{MDM}, we fabricated a nearby reference structure consisting of two identical directional coupler \glspl{MDM} connected by a multimode waveguide with the same length. By comparing the transmission spectra of the inverse-designed \gls{MDM} structure and the reference structure, we extracted the loss and crosstalk attributable to the inverse-designed \gls{MDM} alone. It is worth noting that the directional coupler \gls{MDM} has a length of approximately \SI{800}{\um}, highlighting the substantial footprint reduction achieved by the inverse-designed \gls{MDM}.

The grating coupler used for the TE measurements was a non-uniform design optimized in SPINS-B with a minimum feature size of \SI{400}{nm} \cite{su2020nanophotonic}, achieving an experimental minimum coupling loss of \SI{12.1}{dB} through a single-pitch loopback configuration. The coupling angle for TE measurements was maintained at \ang{30}, aligning with the fiber array. For TM measurements, a regular grating with a period of \SI{1.88}{\um} and a fill factor of 0.5 was used, resulting in an experimental minimum coupling loss of \SI{19.7}{dB}. A \ang{26} coupling angle was employed, increasing the coupling distance to over \SI{100}{\um}. This distance was maintained consistently across all TM measurements to ensure repeatability. The bandwidth of the TM grating coupler was limited to the C-band, as TE crosstalk became significant in the L-band.

\section{Data availability}

The data sets generated during and/or analyzed during this study are available from the corresponding authors on request.

\bibliography{sample}

\section{Acknowledgment}

We thank CMC Microsystems for providing access to the ANT fabrication process. Furthermore, we are grateful to M.Sc. Cameron Horvath from ANT for his help with the description of the fabrication process. We would also like to thank Dr. Anna Wirth-Singh and the customer support team at Ansys Lumerical for the fruitful discussions on topology optimization. This research was funded by the Natural Sciences and Engineering Research Council of Canada (NSERC), the Center for Optics, Photonics and Lasers (COPL), and the Regroupement Stratégique en Microsystèmes du Québec (ReSMiQ).

\section{Author information}

\subsection{Contributions}

J.P. conducted the inverse design optimization of the photonic components. J.P. and N.D. verified the designs and evaluated their robustness against fabrication variations. J.P. prepared the layout and submitted it for fabrication. J.P. and N.D. conceived and designed the experiment. J.P. and N.D. developed and performed all the sequences for the optical measurements. M.M. supervised the project. All authors contributed to discussions, interpretation of the results, and preparation of the manuscript.

\subsection{Corresponding author}

Correspondence to Julian L. Pita Ruiz.

\section{Competing interests}

The authors declare no competing interests

\newpage
\begin{suppinfo}
\section{Supplementary Note 1: Inverse design process}

Supplementary Figures~\ref{fig:S1},~\ref{fig:S2}, and~\ref{fig:S3} present schematic diagrams detailing the main dimensions and initial conditions of the inverse design process, the optimization trajectory, and the material index distribution at representative iterations of each optimization phase for the four-channel \gls{CWDM}, five-mode \gls{MDM}, and \gls{PBS}, respectively. Notably, no significant performance drops are observed during the binarization or \gls{DFM} phases—an issue often encountered in topology optimization processes. This smooth convergence indicates that the fabrication constraints were successfully enforced without disrupting the optimization process.

\renewcommand{\thefigure}{Supplementary Figure \arabic{figure}}
\setcounter{figure}{0}

\renewcommand{\figurename}{Supplementary Figure}
\renewcommand{\thefigure}{\arabic{figure}}
\setcounter{figure}{0}

\begin{figure*}[h!]
\centering
\includegraphics[width = \linewidth]{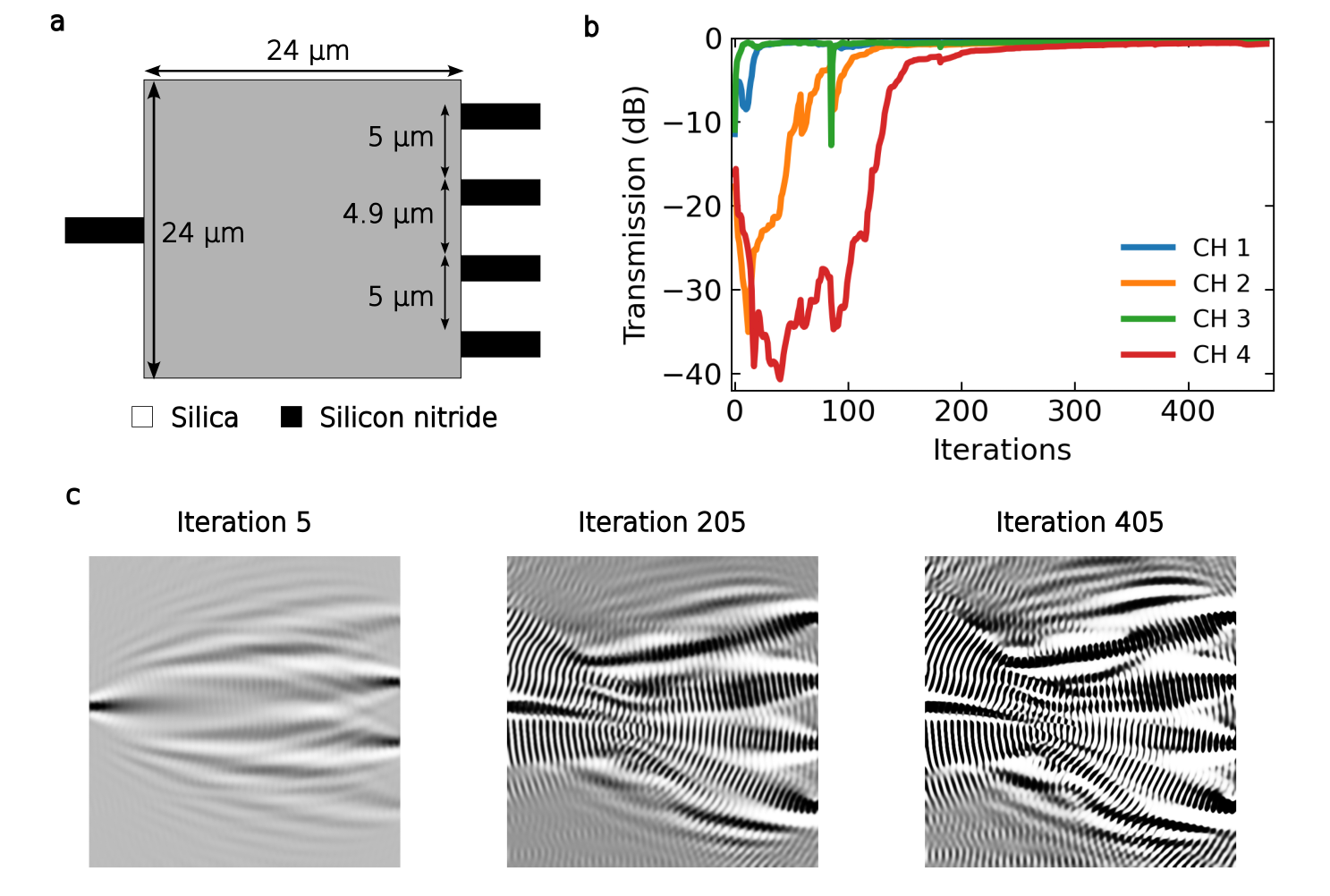}
\caption{Overview of the \gls{CWDM} inverse design process. (a) Device schematic. (b) Optimization trajectory for each channel. (c) Material index distribution at representative iterations for each optimization phase: greyscale, binarization, and \gls{DFM}.}
\label{fig:S1}
\end{figure*}

\begin{figure*}[h!]
\centering
\includegraphics[width = \linewidth]{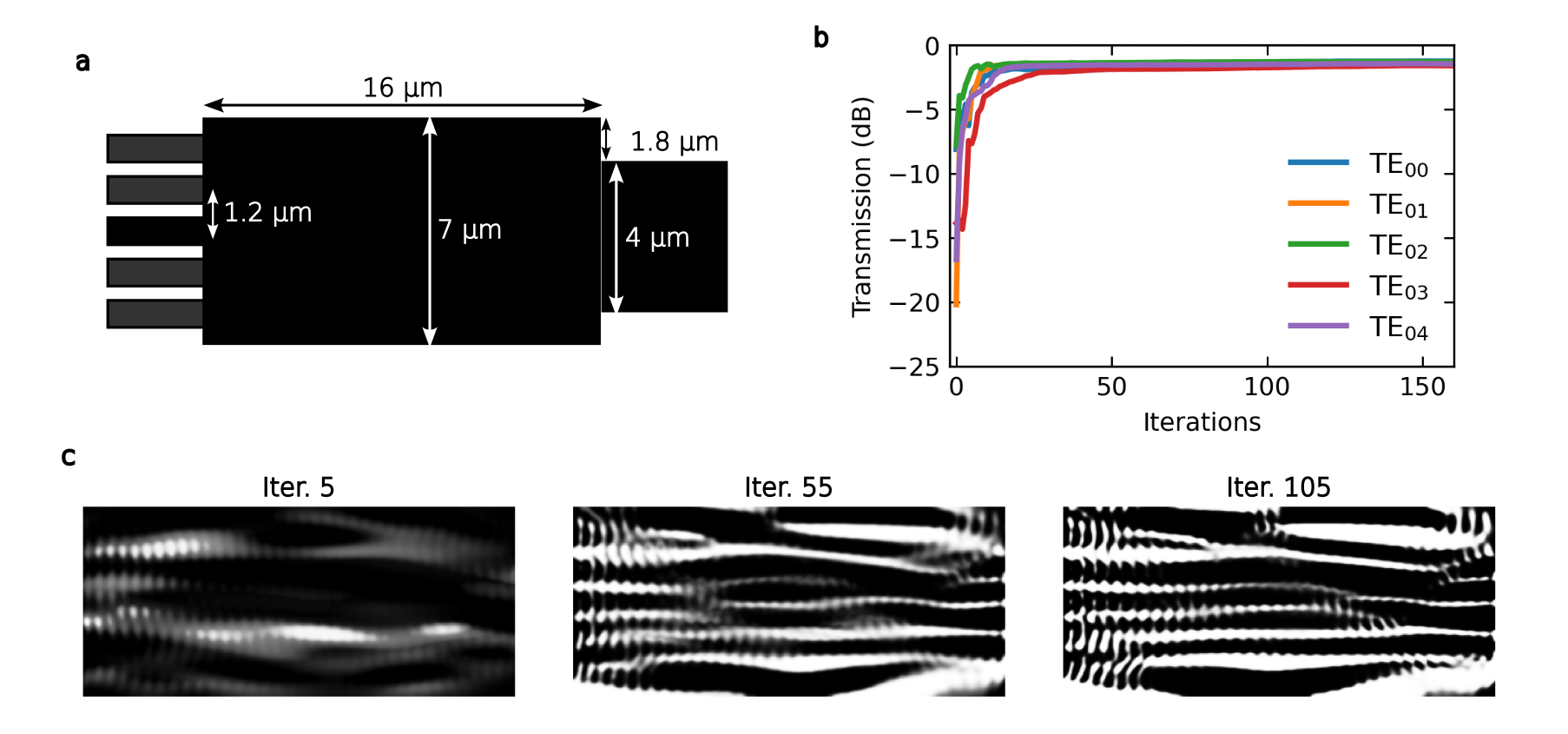}
\caption{Overview of the \gls{MDM} inverse design process. (a) Device schematic. (b) Optimization trajectory for each channel. (c) Material index distribution at representative iterations for each optimization phase: greyscale, binarization, and \gls{DFM}.}
\label{fig:S2}
\end{figure*}

\begin{figure*}[h!]
\centering
\includegraphics[width = \linewidth]{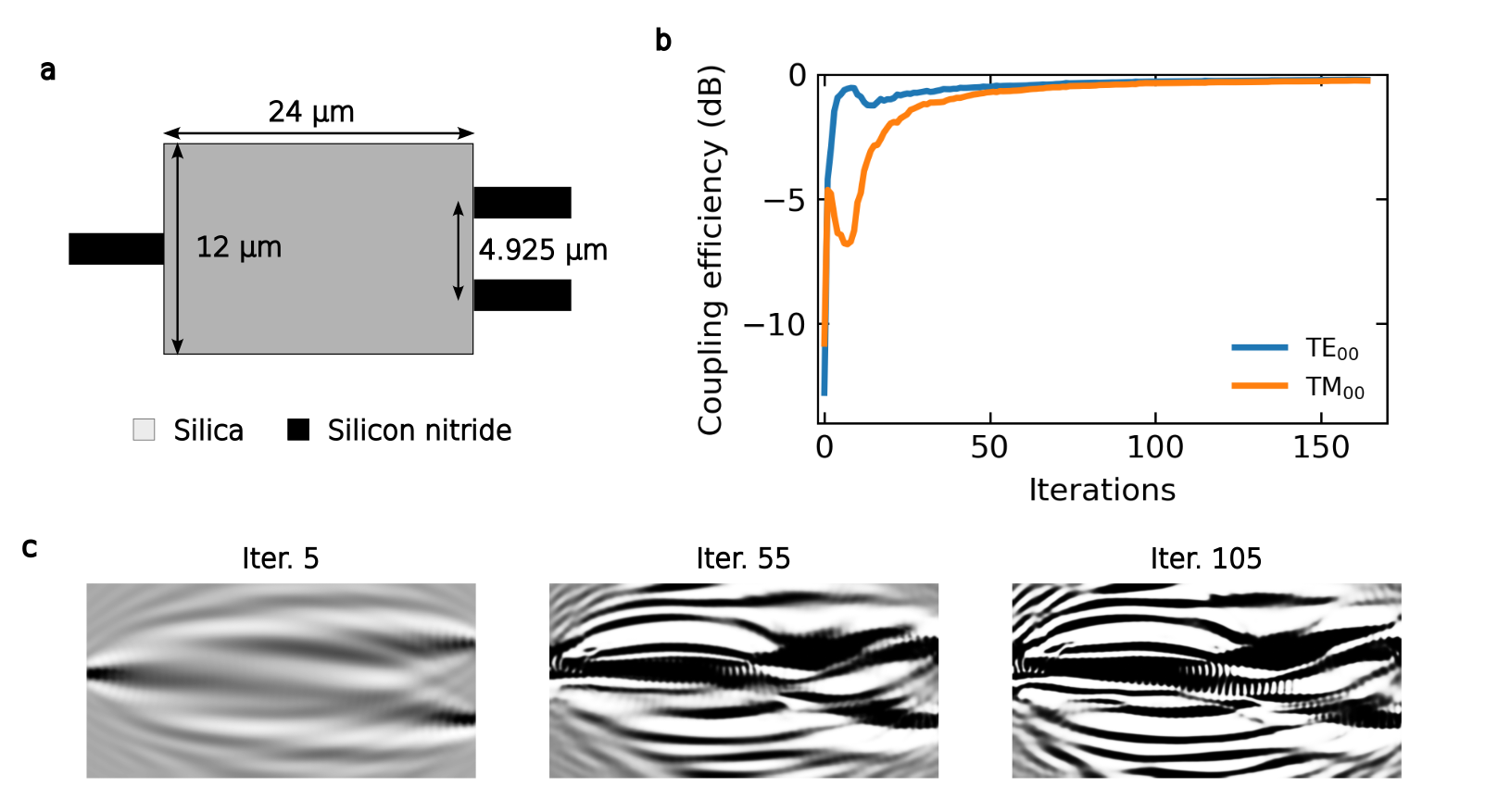}
\caption{Overview of the \gls{PBS} inverse design process. (a) Device schematic. (b) Optimization trajectory for each channel. (c) Material index distribution at representative iterations for each optimization phase: greyscale, binarization, and \gls{DFM}.}
\label{fig:S3}
\end{figure*}
\clearpage
\newpage

\section{Supplementary Note 2: Fabrication variations}

We explore the variations in performance of the \gls{CWDM} and \gls{PBS} devices under fabrication errors using 3D FDTD simulations. For the four-channel \gls{CWDM}, we investigate a lateral over etching of approximately \SI{14}{nm} a $\pm$\SI{1}{\percent} variation in the \ce{SiN} refractive index, whereas for the \gls{PBS}, we examine both lateral under etching and over etching of about \SI{14}{nm}. Supplementary Figure~\ref{fig:S4} presents the simulated transmission and the measured transmission of the over etched design. Additionally, the simulated electromagnetic energy density is shown at \SI{1531}{nm}, \SI{1551}{nm}, \SI{1571}{nm}, and \SI{1591}{nm}. The simulated transmission values were \SI{-1.8}{dB} at \SI{1511.3}{nm}, \SI{-0.9}{dB} at \SI{1531.3}{nm}, \SI{-1.0}{dB} at \SI{1551.3}{nm}, and \SI{-0.9}{dB} at \SI{1571.3}{nm}. The simulated over etched design curves closely match the measurements for each channel, suggesting that the predominant fabrication error in the four-channel \gls{CWDM} device was over etching by approximately \SI{14}{nm}.

\begin{figure*}[h!]
\centering
\includegraphics[width = \linewidth]{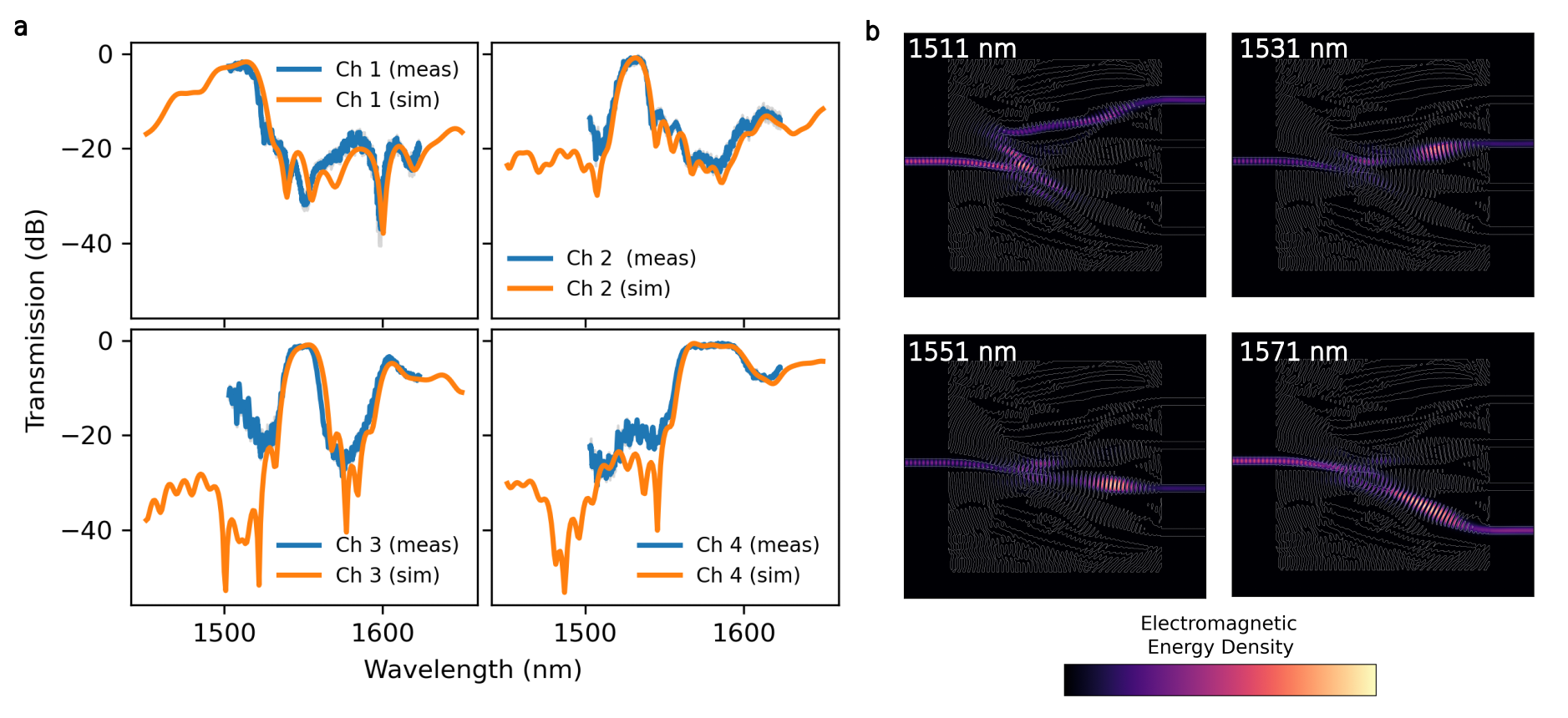}
\caption{Simulation results for all channels of the \gls{CWDM}, considering etching fabrication variations. (a) Simulated transmission efficiency overlaid with the measured transmission efficiency for all four channels. (b) Simulated electromagnetic energy density at \SI{1511}{nm}, \SI{1531}{nm}, \SI{1551}{nm}, and \SI{1571}{nm}.}
\label{fig:S4}
\end{figure*}

Supplementary Figure~\ref{fig:S4b} shows the behavior of the four-channel \gls{CWDM} under variations in the \gls{SiN} core refractive index of $\pm$\SI{1}{\percent}. When the index increases by \SI{1}{\percent}, all channels experience a shift of approximately \SI{10}{\nano\meter} toward longer wavelengths, with transmission efficiencies of \SI{-0.6}{\decibel}, \SI{-0.4}{\decibel}, \SI{-0.1}{\decibel}, and \SI{-0.4}{\decibel} at \SI{1540}{\nano\meter}, \SI{1560}{\nano\meter}, \SI{1580}{\nano\meter}, and \SI{1600}{\nano\meter}, respectively. The worst crosstalk occurs in channel 3, reaching \SI{-14.1}{\decibel}. When the index decreases by \SI{1}{\percent}, all channels shift by approximately \SI{8}{\nano\meter} toward shorter wavelengths. The corresponding transmission efficiencies are \SI{-0.8}{\decibel}, \SI{-0.4}{\decibel}, \SI{-0.3}{\decibel}, and \SI{-0.5}{\decibel} at \SI{1522}{\nano\meter}, \SI{1542}{\nano\meter}, \SI{1562}{\nano\meter}, and \SI{1582}{\nano\meter}, respectively. The worst crosstalk again occurs in channel 3, measured at \SI{-14.9}{\decibel}. These results indicate that changes in refractive index shift the central wavelengths and slightly improve peak transmission efficiency at the shifted wavelengths. Crosstalk also improves by at least \SI{0.8}{\decibel} in the case of the $+$\SI{1}{\percent} variation.

\begin{figure}[h!]
\centering
\includegraphics[scale=1.1]{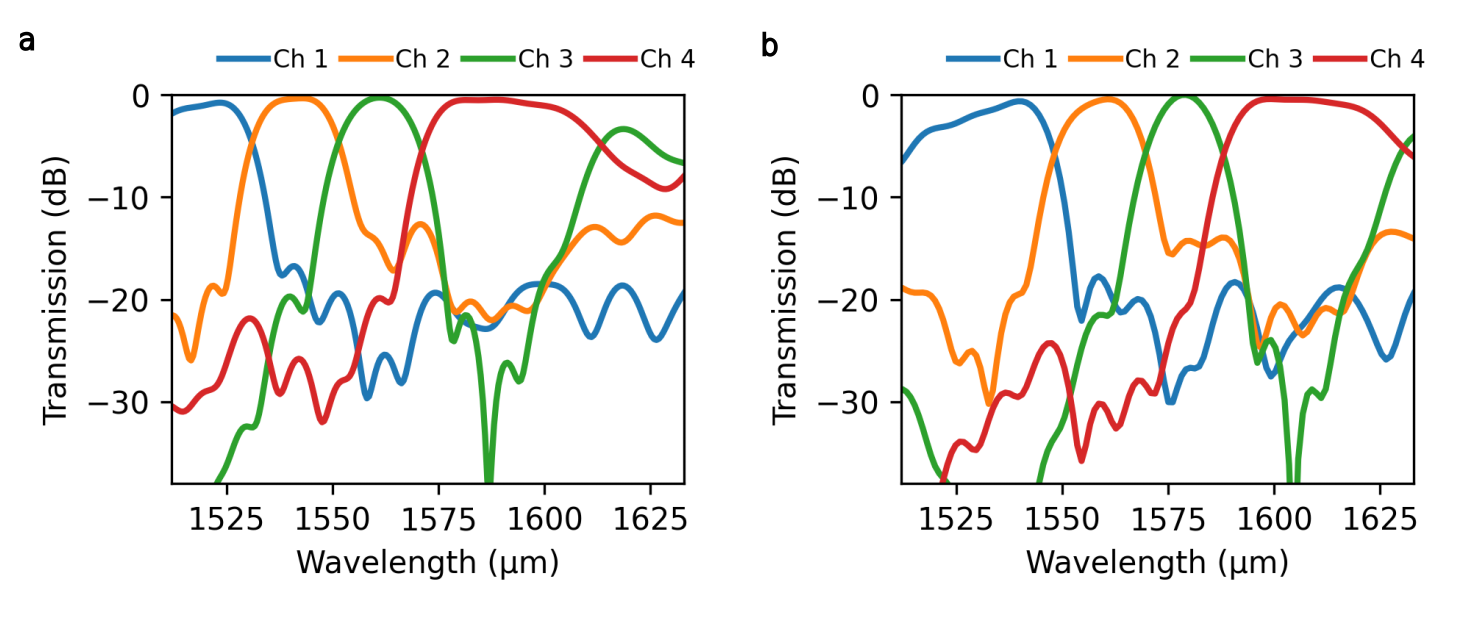}
\caption{Simulated transmission efficiency of the four-channel \gls{CWDM}, considering \gls{SiN} refractive index variations. (a) Simulated transmission calculated using 3D \gls{FDTD} with a \gls{SiN} refractive index variation of \SI{+1}{\percent}. (b) Simulated transmission for a \gls{SiN} refractive index variation of \SI{-1}{\percent}.}
\label{fig:S4b}
\end{figure}

Supplementary Figure~\ref{fig:S5} presents the simulated efficiency of the \gls{PBS} under two fabrication variations (FVs) to assess its robustness against fabrication deviations. FV1 corresponds to a lateral over etching of approximately \SI{14}{nm}, whereas FV2 represents a lateral under etching of the same magnitude. For FV1, the average simulated TE transmission of channel 1 is \SI{-0.6}{dB}, with an undesired TM transmission averaging \SI{-17.2}{dB} across the C-band. For channel 2, the average simulated TM transmission is \SI{-0.5}{dB}, whereas the undesired TE transmission averages \SI{-21.3}{dB}. For FV2, the average simulated TE transmission of channel 1 is \SI{-0.4}{dB}, with the undesired TM transmission averaging \SI{-23.5}{dB}. Similarly, the average simulated TM transmission of channel 2 is \SI{-0.3}{dB}, with the undesired TE transmission averaging \SI{-22.3}{dB}. Comparing both cases, FV1 (lateral over etching) has a more pronounced impact on the \gls{PBS} performance, particularly increasing polarization crosstalk. Moreover, FV1 shows a stronger correlation with the experimental measurements, suggesting that over etching was the dominant fabrication error in the manufactured devices.

\begin{figure*}[h!]
\centering
\includegraphics[width = \linewidth]{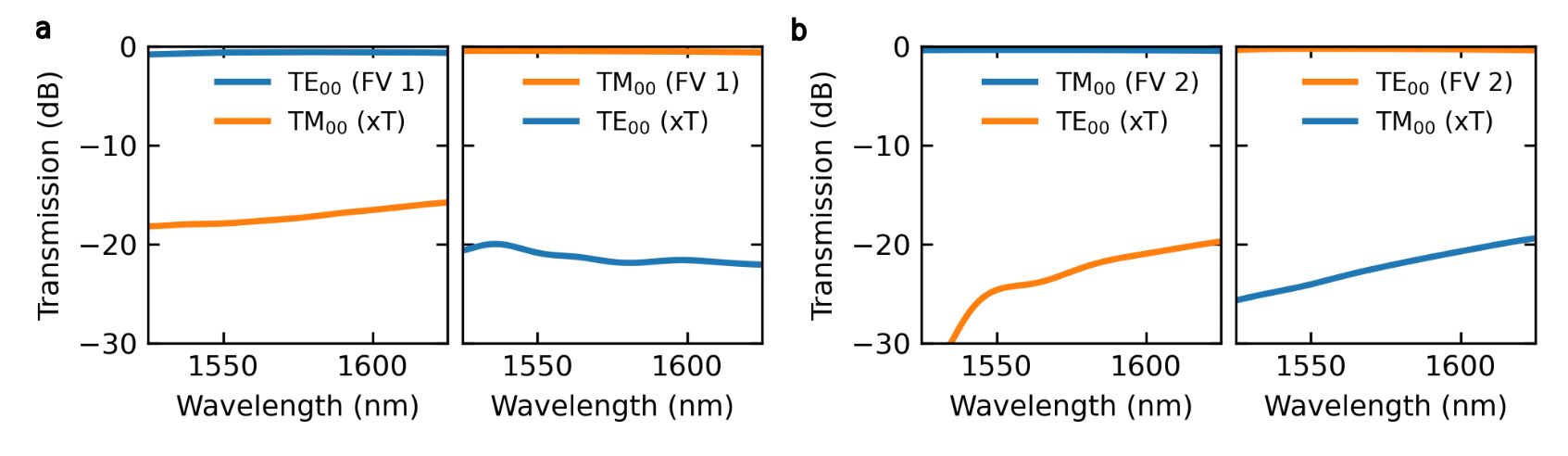}
\caption{Simulation results for both channels and polarizations of the \gls{PBS}, considering fabrication variations. (a) FV1: over etching, and (b) FV2: under etching.}
\label{fig:S5}
\end{figure*}

\section{Supplementary Note 3: \gls{CWDM} Movie}

The supplementary movie shows the electromagnetic energy density distribution in a horizontal cross-section through the center of the four-channel \gls{CWDM} as a function of wavelength.

\newpage
\section{Supplementary Note 4: Topology Optimization Overview}

In this section we review topology optimization formulation based on the following references [1-6]. First, we formulate the electromagnetic topology optimization problem, where the underlying physics are governed by Maxwell's equations, which, for time-harmonic fields, reduce to the Helmholtz equation.

\begin{equation}
\begin{array}{ll}
\underset{\boldsymbol{\rho}}{\text{minimize}} &  f(\mathbf{E}(\boldsymbol{\rho}))   \\
\text{subject to} & \nabla \times \left(\frac{1}{\mu_0\mu_r} \nabla \times \mathbf{E} \right) - \omega_m^2 \epsilon_0 \epsilon_r(\boldsymbol{\rho}) \mathbf{E} = -i\omega_m \mathbf{J}, \quad m \in \{1, 2, \dots, M\} \\
& 0 \le \boldsymbol{\rho} \le 1 \\
& g_k(\boldsymbol{\rho}) \le 0, \quad k \in \{1, 2, \dots, K\}
\end{array}
\end{equation}

Here, $f$ is the objective function to be minimized and depends on the electric field $\mathbf{E}$, which is the system state variable determined by solving the governing equation for each excitation frequency $\omega_m$ and current density source $\mathbf{J}$. The material properties—specifically the relative permittivity $\epsilon_r(\boldsymbol{\rho})$—are determined by the material density field $\boldsymbol{\rho}$ (the design variable, constrained between 0 and 1), and $g_k(\boldsymbol{\rho})$ represents additional inequality constraints on the material distribution.   

LumOpt uses a three-field topology optimization scheme that utilizes a design parameters field $\boldsymbol{\rho}$, a filtered field $\tilde{\boldsymbol{\rho}}$ and a physical field $\bar{\boldsymbol{\rho}}$. Their relations are established through the following filtering and threshold processes.

\begin{equation}
\tilde{\boldsymbol{\rho}} = k(\mathbf{x}) * \boldsymbol{\rho}
\end{equation}

Here, $\tilde{\boldsymbol{\rho}}$ is the filtered design field, $*$ denotes a 2D convolution and $k(\mathbf{x})$ is a circular top-hat kernel with radius $R$ centered at $\mathbf{x}_0$:

\begin{equation}
k(\mathbf{x}) = \begin{cases}
\frac{1}{\pi R^2}, & \text{if } \|\mathbf{x} - \mathbf{x}_0\| \le R \\
0, & \text{otherwise}
\end{cases}
\end{equation}

Next, the filtered design field $\tilde{\boldsymbol{\rho}}$ is projected to a nearly binary physical design using a differentiable approximation of the Heaviside function:

\begin{equation}
\bar{\boldsymbol{\rho}} = \frac{\tanh(\beta \eta) + \tanh(\beta (\tilde{\boldsymbol{\rho}} - \eta))}{\tanh(\beta \eta) + \tanh(\beta (1 - \eta))}
\end{equation}

Here, $\beta$ controls the sharpness of the transition, and $\eta$ sets the threshold around which projection occurs.

The final relative permittivity is interpolated as:
\begin{equation}
\epsilon_r(\bar{\boldsymbol{\rho}}) = \epsilon_{\text{min}} + \bar{\boldsymbol{\rho}} (\epsilon_{\text{max}} - \epsilon_{\text{min}}),
\end{equation}

where $\epsilon_{\text{min}}$ and $\epsilon_{\text{max}}$ correspond to the permittivities of the cladding and core materials, respectively.

To ensure manufacturability, in the DFM phase two geometric constraints are imposed to enforce a minimum length scale on $\bar{\rho}_\eta$, thresholded within the range $\eta \in (\eta_d, \eta_e)$. These constraints are defined as follows:
\begin{align}
g^s &= \frac{1}{n} \sum_{i \in \mathbb{N}} I_i^s \cdot \left[\min(\tilde{\rho}_i - \eta_e, 0)\right]^2 = 0 \\
g^v &= \frac{1}{n} \sum_{i \in \mathbb{N}} I_i^v \cdot \left[\min(\eta_d - \tilde{\rho}_i, 0)\right]^2 = 0,
\end{align}

where $n$ is the number of discretized elements in the domain $\mathbb{N}$. These constraints ensure that the filtered design field $\tilde{\rho}_i$ is larger than $\eta_e$ in the inflection region of the solid phase and smaller than $\eta_d$ in the inflection region of the void phase, thereby satisfying the minimum length scale requirement. To identify the inflection regions, the following structural indicator functions are defined as:
\begin{align}
I^s &= \bar{\rho} \cdot \exp\left(-c \cdot |\nabla \tilde{\rho}|^2 \right) \\
I^v &= (1 - \bar{\rho}) \cdot \exp\left(-c \cdot |\nabla \tilde{\rho}|^2 \right)
\end{align}

$I^s$ and $I^v$ identify the inflection regions of the solid phase and void phase, respectively. The parameter $c$ in the exponential term controls how sharply these inflection regions are defined.

\section*{Supplementary References}

\begin{enumerate}
\item Wang, F.; Jensen, J. S.; Sigmund, O. Robust topology optimization of photonic crystal waveguides with tailored dispersion properties. \textit{J. Opt. Soc. Am. B} \textbf{2011}, \textit{28} (3), 387-397.

\item Jensen, J. S.; Sigmund, O. Topology optimization for nano-photonics. \textit{Laser Photonics Rev.} \textbf{2011}, \textit{5} (2), 308-321.

\item Sigmund, O. Topology optimization: a tool for the tailoring of structures and materials. \textit{Philos. Trans. R. Soc. A} \textbf{2013}, \textit{371} (1993), 20120462.

\item Zhou, M.; Lazarov, B. S.; Wang, F.; Sigmund, O. Minimum length scale in topology optimization by geometric constraints. \textit{Comput. Methods Appl. Mech. Eng.} \textbf{2015}, \textit{293}, 266-282.

\item Christiansen, R. E.; Sigmund, O. Inverse design in photonics by topology optimization: tutorial. \textit{J. Opt. Soc. Am. B} \textbf{2021}, \textit{38} (2), 496-509.

\item Hammond, A. M.; Camacho, R. M.; Oskooi, A. Photonic topology optimization with semiconductor-foundry design-rule constraints. \textit{Opt. Express} \textbf{2021}, \textit{29} (15), 23916-23938.
\end{enumerate}
\end{suppinfo}
\end{document}